\documentclass[twocolumn,twoside]{revtex4}
\usepackage{graphicx}
\usepackage{fancyhdr}
\begin{document}

\title{CP Violation in Charmed and Bottom Baryon Physics}

%

\author{John~M.~Yelton}
\affiliation{University of Florida, Gainesville, FL, USA, 32611}

\begin{abstract}
  This is an experimentalist's view of the recent results on, and prospects
  for, CP Violation in charmed and bottom baryons.
\end{abstract}

\maketitle

\thispagestyle{fancy}


\section{Introduction}

The phenomenon of CP violation has been discussed in several contributions to this
conference already. However, these studies have focussed on mesons. Studies of baryon decays
are usually one or decades behind those of mesons, but in CP violation the situation is even 
more difficult. Whereas in neutral mesons there can be mixing between particle and 
anti-particle, and two distinct interfering amplitudes which can produce time-dependent
CP violation, baryon number is a very strong constraint and the window provided by mixing
is not available. This means that any CP violation 
in baryons needs to be $direct$. However, though the difficulties are many, the prize is great. 
If we want to understand why we are here today (and made out of matter), we need to understand
CP violation in baryons. 

This is an experimentalist's review of recent results on, and prospects for, CP violation 
studies in charmed and bottom baryons.

\section{The Experiments}

It is no surprise to find that the two featured current experiments are LHCb (representing the ``energy
frontier'' of $pp$ collisions), and Belle~II (representing the ``intensity frontier'' of
$e^+e^-$ annihilation).

Table~\ref{tab:comparison} shows the comparison between the two. The first and last rows show
huge benefits of LHCb. These are partly compensated by the other entries in the table, which 
means that there will be a complementarity of approaches in the next few years. 

\begin{table}[h]
\begin{center}
\caption{Comparison of LHCb and Belle~II for CP violation studies in baryons.}
\begin{tabular}{c|c|c}
\hline
$\sigma(b+\overline{b}){\rm nb}$ & $\sim 150,000$ $\sim 1$ \\
\hline 
$\int Ldt({\rm fb}^{-1})$  & $\sim 50$ & $\sim 50,000$ (eventually...) \\
\hline 
Background Level & Higher & Lower \\
\hline
Typical Efficiency & Lower & Higher \\
\hline 
Neutral Efficiency & Lower & Higher \\
\hline
Decay time resolution & Great & Good \\
\hline
Production Ratios & Baryons produced & Baryons/Antibaryons \\
& preferentially & produced equally \\
\hline
Collision spot size & Bigger & Tiny \\
\hline
Particles produced & Includes b baryons & Only $c$ and $s$ \\
\end{tabular}
\label{tab:comparison}
\end{center}
\end{table}

\section{LHC\MakeLowercase{b} Search for CP Violation in $\Xi_b\to pK^-K^-$}

Why look at this mode? Interference between two amplitudes 
with different weak and strong phases leads 
to CP violation in decay.
Weak phases are associated with the complex elements of the CKM matrix 
Strong phases associated with hadronic final-state effects. 
In general, the common decays of $b$ hadrons
to charm quarks have one amplitude that dominates and so there is 
no useful interference between amplitudes
to use for CP violation studies. This is similar to the situation in meson decays.

This analysis is presented in~\cite{XIB}. The signal is around 500 events. 
As CP violation is expected to be
enhanced in the baryon resonance region, the analysis performs a fit to the 
entire Dalitz plot taking
into account the six excited $\Lambda$ and $\Sigma$ resonances 
tabulated in the Particle Data Book~\cite{PDG} 
in the region 1.4-2.0 ${\rm GeV}/c^2$, 
searching for a significant difference between particle and anti-particles. 
The results are that there is no significant CP violation in any of 
the six quasi-two-body decays. 
The statistical uncertainties are large (not surprising when several resonances
are barely significant), and so will improve with more data. 
The systematic uncertainties will also improve.

\section{LHC\MakeLowercase{b} Search for CP Violation in $\Lambda_b\to DpK^-$}

This is suppressed decay mode and comes in different varieties. The one
that is interesting 
is $\Lambda_c^0\to[K^+\pi^-]_D pK^-$ where the kaons have different charges.
Note that we do not specify whether the $K\pi$ is a $D$ or a $\bar{D}$ as
both can contribute. This decay has contributions both from $b \to c$ 
and $b\to u$ which are of the same order of magnitude, and an interference
that depends on the CKM angle $\gamma$. 

What is measured is the asymmetry:

$A = \frac{(B_{particle} - B_{antiparticle})}
{(B_{particle} + B_{antiparticle})}$ ,

\noindent{
where $B_{particle}= B(\Lambda_b^0\to[K^+\pi^-]_D pK^-)$
and 
$B_{antiparticle}= 
B({\bar \Lambda}_b^0\to[K^-\pi^+]_{D} {\bar p}K^+)$.
Note that the interference is anticipated to be larger 
in regions of phase space that involve excited $\Lambda$ states.}

The analysis is presented in~\cite{DpK}. The measured CP asymmetries are:

\centerline{$A_{CP}=0.12\pm0.09{\rm (stat)}^{+0.02}_{-0.03}({\rm syst.})$ }

and

\centerline{$A_{CP}=0.01\pm0.16{\rm (stat)}^{+0.02}_{-0.03}({\rm syst.})$ }

\noindent{where this first is integrated over all phase-space and the second in the low-mass
(resonance) $pK$ region. The conclusion is that there is no evidence of CP violation. }

\section{LHC\MakeLowercase{b} Search for CP Violation in $\Lambda_b^0\to D\pi^-\pi^+\pi^-$}

There are two methods now in use to analyze this decay mode, which has the benefit
of relatively high statistics~\cite{p3pi}. The first uses ``scalar triple products'', defined
in~\cite{GD}, which involve the momentum correlation between the proton, the positive pion
and the ``faster'' of the negative pions, and the analogs for antiparticles.

$ C = {\vec p_p} \cdot  ({\vec p}_{\pi^-} \times {\vec p}_{\pi^+})$
and
${\bar  C} = {\vec p_{\bar p}} \cdot  ({\vec p}_{\pi^+} \times {\vec p}_{\pi^-})$

\noindent{
The data is then divided into four pieces, each with yields N (i.e. divide by two by
particle/antiparticle, and again by 2 according to the sign of the scalar triple product.

The Triple Product Asymmetry is then defined for particles and antiparticle:
}

$A=\frac{N(C_T > 0) - N(C_T < 0)}{N(C_T > 0) + N(C_T < 0)}$
and
${\bar A}=\frac{{\vec N}({\vec C_T} > 0) -  {\bar N}({\bar C_T} < 0)}
  {{\bar N}({\bar C_T} > 0) + {\bar N}({\bar C_T} < 0)}$,
  
  \noindent{  where these two would be the same if there is no CP violation. Then, the CP violating piece looked for is defined as:}

  $a_{CP} = \frac{1}{2}(A_T-{\bar A_T})$

{\noindent Integrated over all phase-space they find
$a_{CP} = (-0.7\pm0.7\pm0.2)\%$
which is clearly consistent with zero.}

Once again there is the expectation that any CP Violation will
be in any area of phase-space dominated by resonances, and here
there are two sorts of resonances - excited nucleons ($N^*$), and
excited mesons (e.g. $a_1$). Each of these preferred regions
are then divided
into bins dependent on the relative angles of the final state particles.

The biggest discrepency from the non-CPV expectation is found
in the $N^*$ enhanced sample when plotted in terms of the angle
between the decay planes of the $(\pi^+\pi^-)_{slow}$ and
$(p\pi^-)_{fast}$. This discrepency is of 2.9 $\sigma$ significance. A previous
version of this analysis provided some excitement as it found
3.3 $\sigma$ significance on 25\% of the dataset, but the extra
data has not enhanced the effect.

LHCb have also analyzed the data in a less model dependent manner
using independent unbinned tests~\cite{MW}. Again, no significant evidence
of CP violation has been found. 

\section{CP Violation Searches using Charmed Baryons}

The charm sector has some experimental advantages over the B sector.
One is the complementary datasets available in $e^+e^-$
annihalations, and the other is the comparatively low multiplicies
in the final states.

The basic research thrusts comprise choosing a suitable decay mode
and measuring its branching fraction, measuring the asymmetry parameter,
$\alpha$ defined below, and then measuring the difference in asymmetry paramaters
in particles and anti-particles. Once again, no measurable CPV is
expected in 2-body Cabibbo-favored decays, and singly-Cabibbo-suppressed
decays are a more promising laboratory, with effects of order $10^{-3}$
to be expected.

Typical decay modes that we can hope to produce interesting results
are $\Lambda_c^+\to\Lambda K^+$ and $\Lambda_c^+\to\Sigma^0K^+$ 
which have measured branching fractions of $(6.1\pm1.2)\times 10^{-4}$
and $(5.2\pm 0.8)\times 10^{-4}$ respectively. In $e^+e^-$ machines
these (particularly the former) can be detected with good efficiency
and purity.

In decays of $\Lambda_c^+$ to a baryon and pseudo-scalar meson, the $\alpha$
parameter is defined to be: $\alpha = \frac{2Re(s \cdot p}{|s|^2 + |p|^2}$
where $s$ and $p$ are the parity-violating s-wave and the parity-conserving
p-wave amplitudes. Operationally, for a decay such as
$\Lambda_c^+\to\Lambda K^+$, we define an angle $\theta_{\Lambda}$, which is the angle between the proton momentum in the $\Lambda$ frame and the $\Lambda$
in the $\Lambda_c^+$ frame.
The data is divided into bins of $cos(\Theta_{\Lambda})$, and then use the
equation:

$dN/dcos \theta_{\Lambda}\propto 
1 + \alpha(\Lambda_c^+ \to \Lambda K^+)\alpha
(\Lambda\to p\pi^-)cos\theta_{\Lambda}$

In other words, split the data into bins of $cos\theta_{\Lambda}$,
fit the mass peak in each plot, find the slope and extract the product
of the $\alpha$ parameters ($\alpha$ for a $\Lambda$ is well-known).

The $\alpha$ parameters are found separately for particles and antiparticles, and
the then the CP-violating parameter defined as:

\centerline{
$A_{CP}^{\alpha} = \frac{\alpha(\Lambda_c^+)+\alpha(\Lambda_c^+)}
{\alpha(\Lambda_c^+)-\alpha(\Lambda_c^+)}$
}

The $\alpha$ parameters are interesting in their own right, as they are parameters,
along with the partial width, that can in principle be predicted theoretically
in different models of weak decays. However, until recently, neither the experimental
measurements or the theoretical models were of sufficient quality for such 
comparisons to be meaningful. However, the situation is changing. For example,
in 2001 CLEO measured $\alpha$ for $\Xi_c^0\to\Xi^-\pi^+$ to be 
$-0.56\pm 0.39^{+0.10}_{-0.09}$, 
but that paper
did not receive a single citation for the first eight years! Recently, BELLE~\cite{BXic} measured
$\alpha$ for particles and anti-particles and found $A_{CP}^{\alpha} = 0.024\pm 0.052\pm 0.012$.
This, of course is a Cabibbo-allowed decay and finding CP violation 
in such a mode would be a major surprise.
However, in the Cabibbo-suppressed modes $\Lambda_c^+\to\Lambda K^+$ and 
$\Lambda_c^+\to \Sigma^0K^+$ precisions of the order of 0.1 and 0.3 would seem to be
attainable for each $1\ ab^{-1}$ of data (i.e. the already taken Belle dataset). Although
this may not be sufficient, the prospect of $50\ {\rm ab}^{-1}$ Belle~II data opens
realistic possibilities for these and other similar modes.

\section{ LHC\MakeLowercase{b} Measurements in $\Lambda_c^+\to ph^+ h^-$}

These, with $h = \pi$ or $K$, are two Cabibbo-suppressed decays of the $\Lambda_c^+$, 
and the big challenge is to 
understand the detection efficiency of protons and anti-protons. One way of avoiding 
this~\cite{LHCphh}
is to use two related decays and compare them. That is, instead of

$A_{CP}(f) = \frac{\Gamma(f)-\Gamma(\bar{f})} 
{\Gamma(f)+\Gamma(\bar{f})}$

use

$A_{Raw} = \frac{N(f)-N(\bar{f})} 
{N(f)+N(\bar{f})}$

\noindent{where $f$ refers to a specific final state and $N$ is the 
number detected.}

Then we can say that the difference in the $A_{Raw}$ parameters
will be a good approximation to the the difference of the $A_{CP}$
parameters. 

Although the raw asymmetries of the two modes are significantly non-zero, 
the difference is $\Delta_{CP} = (0.30\pm0.91\pm0.61)\%$ and therefore there
is no evidence of CP violation.

\section{ LHC\MakeLowercase{b} Measurements in $\Xi_c^+ \to K^+ \pi^-$}

This Cabibbo-suppressed decay is a very useful one for LHCb as the
statistics are very high and the background suppression good~\cite{pKpi}. 
The Cabibbo-allowed decay of $\Lambda_c^+\to pK^-\pi^+$ is used as a control. 
The substructure of these decays are very rich, and CP violation, if it exists
in these modes, is likely to depend on where in the Dalitz plot the decay lies.
Two different analysis techniques are used. One is binned into
areas of the Dalitz plot, searching for localized differences 
in the plots for baryons and anti-baryons. The second is unbinned.
However, no significant differences are found between baryons
and anti-baryons, and the results are consistent with the hypothesis of
no CP violation.

\section{ Conclusions}

I have reviewed a number of searches for CP violation in charmed baryon 
baryons. No significant violation has been observed. 
However, the sensitivity is not at the level that this is a surprise.

It is clear that more predictions for specific decay modes for CPV
would be very useful. I expect in the next few years to have more
measurements of the asymmetry parameter, $\alpha$, whch are 
interesting in their own right and can lead to measurements
of CP violation. I would like to point out that many of the 
searches involve knowledge of the excited hyperons in the 
substructure of the heavy baryon decays. Research 
in the area of excited hyperons is under-appreciated. However, clearly
the biggest challenge in the next decade is to collect big enough
datasets that comparatively rare decays can be studied with 
good statistics.

\begin{acknowledgments}
  This work is supported in part be DOE contract DE-SC0009824.
  
\end{acknowledgments}

\bigskip 

\end{document}